\documentclass[runningheads]{llncs}
\usepackage{graphicx}

\usepackage{enumitem}
\usepackage[utf8]{inputenc}
\usepackage{color}
\usepackage{soul}
\usepackage{graphicx}
\usepackage{xspace}
\usepackage{subcaption}
\usepackage{float}
\usepackage{multirow}
\usepackage{flushend}

\usepackage[ruled,noend]{algorithm2e}

\usepackage{libertine}
\usepackage{libertinust1math}
\usepackage[T1]{fontenc}

\everypar{\looseness=-1}

\let\OLDthebibliography\thebibliography
\renewcommand\thebibliography[1]{
	\OLDthebibliography{#1}
	\setlength{\parskip}{0pt}
	\setlength{\itemsep}{0pt plus 0.3ex}
}

\usepackage{pbox}
\begin{document}

\title{One Table to Count Them All:\\Parallel Frequency Estimation on Single-Board Computers}

\titlerunning{Parallel Frequency Estimation on Single-Board Computers}
\author{Fatih Ta\c{s}yaran \and
Kerem Y{\i}ld{\i}r{\i}r\and \\ 
Kamer Kaya \and Mustafa Kemal Ta\c{s} }
\authorrunning{Ta\c{s}yaran et al.}
\institute{Sabanc{\i} University, Istanbul, Turkey, \\ \email{\scriptsize{\{fatihtasyaran,keremyildirir,kaya,mkemaltas\}@sabanciuniv.edu}}}

\maketitle            
\begin{abstract}
\begin{sloppypar}
Sketches are probabilistic data structures that can provide approximate results within mathematically proven error bounds while using orders of magnitude less memory than traditional approaches. They are tailored for streaming data analysis on architectures even with limited memory such as single-board computers that are widely exploited for IoT and edge computing. Since these devices offer multiple cores, with efficient parallel sketching schemes, they are able to manage high volumes of data streams. However, since their caches are relatively small, a careful parallelization is required.

In this work, we focus on the frequency estimation problem and evaluate the performance of a high-end server, a 4-core Raspberry Pi and an 8-core Odroid. As a sketch, we employed the widely used Count-Min Sketch. 
To hash the stream in parallel and in a cache-friendly way, we applied a novel tabulation approach and rearranged the auxiliary tables into a single one. To parallelize the process with performance, we modified the workflow and applied a form of buffering between hash computations and sketch updates. 

Today, many single-board computers have heterogeneous processors in which slow and fast cores are equipped together. To utilize all these cores to their full potential, we proposed a dynamic load-balancing mechanism which significantly increased the performance of frequency estimation. 

\vspace{3mm}
\textbf{Keywords: } Parallel algorithms $\cdot$ streaming data $\cdot$ single board computers
\end{sloppypar}
\end{abstract}

\section{Introduction}

Although querying streaming data with 100$\%$ accuracy may be possible by using cutting edge servers equipped with a large memory and powerful processor(s), enabling power efficient devices such as single-board computers~(SBCs), e.g., Arduino, Raspberry Pi, Odroid, with smarter algorithms and data structures yields cheaper and energy efficient solutions. These devices are indeed cheap, are equipped with multicore processors, and portable enough to be located at the edge of a data ecosystem, which is where the data is actually generated. Furthermore, SBCs can be enhanced with various hardware such as cameras, sensors, and software such as network sniffers. Hence, exploiting their superior price/performance ratio for data streams is a promising approach. A comprehensive survey of data stream applications can be found in~\cite{muthukrishnan2005}. 
      
Sketches can be defined as data summaries and there exist various sketches in the literature tailored for different applications. These structures help us process a query on a massive dataset with small, usually sub-linear amount of memory~\cite{alon1996,charikar2002,dobra2002,gilbert2002}. Furthermore, each data stream can be independently sketched and then these sketches can be combined to obtain the final sketch. Due to the implicit compression, there is almost always a trade-off between the accuracy of the final result and the sketch size. A complete analysis and comparison of various sketches can be found in~\cite{cormode2005}.

{\it Count-Min Sketch}~(CMS) is a probabilistic sketch that helps to estimate the frequencies, i.e., the number of occurrences, of the items in a stream~\cite{cormode2005}. The frequency  information is crucial to find heavy-hitters or rare items and detecting anomalies~\cite{cormode2003,cormode2005}. A CMS stores a small counter table to keep the track of the frequencies. The accesses to the sketch are decided based on the hashes of the items and the corresponding counters are incremented. Intuitively, the frequencies of the items are not exact due to the hash collisions. An important property of a CMS is that the error is always one sided; that is, the sketch never underestimates the frequencies. 

Since independent sketches can be combined, even for a single data stream, generating a sketch in parallel is considered to be a straightforward task; each processor can independently consume a different part of a stream and build a partial sketch. However, with $\tau$ threads, this straightforward approach uses $\tau$ times more memory. Although this may not a problem for a high-end server, when the cache sizes are small, using more memory can be an important burden. In this work, we focus on the frequency estimation problem on single-board multicore computers. Our contributions can be summarized as follows:

\vspace{-0.2\topsep}
\begin{enumerate}[leftmargin=*]
\item We propose a parallel algorithm to generate a CMS and evaluate its performance on a high-end server and two multicore SBCs; Raspberry Pi 3 Model B+ and Odroid-XU4.  We restructure the sketch construction phase while avoiding possible race-conditions on a {\em single} CMS table. With a single table, a careful synchronization is necessary, since race-conditions not only degrade the performance but also increase the amount of error on estimation. Although we use CMS in this work, the techniques proposed in this paper can easily be extended to other table-based frequency estimation sketches such as Count-Sketch and Count Min-Min Sketch. 

\item Today, many SBCs have fast and slow cores to reduce the energy consumption.  However, the performance difference of these heterogenous cores differ for different devices. Under this heterogeneity, a manual optimization is required for each SBC. As our second contribution, we propose a load-balancing mechanism that distributes the work evenly to all the available cores and uses them as efficient as possible. The proposed CMS generation technique is dynamic; it is not specialized for a single device and can be employed on various devices having heterogeneous cores. 

\item 
As the hashing function, we use {\em tabulation hashing} which is recently proven to provide strong statistical guarantees~\cite{thorup2017} and faster than many hashing algorithms available; a recent comparison can be found in~\cite{Dahlgaard2017}. For some sketches including CMS, to reduce the estimation error, the same item is hashed multiple times with a different function from the same family. As our final contribution, we propose a cache-friendly tabulation scheme to compute multiple hashes at a time. The scheme can also be used for other applications using multiple tabulation hashes.
\end{enumerate}
\vspace{-0.2\topsep}

 \section{Notation and Background}\label{sec:not}

Let $\mathcal{U} = \{1,\cdots,n\}$ be the universal set where the elements in the stream are coming from. 
Let $N$ be size of the stream ${\tt s}[.]$ where ${\tt s}[i]$ denotes the $i$th element in the stream.
We will use $f_x$ to denote the frequency of an item. Hence, $$f_x = |\{x = {\tt s}[i]: 1 \leq i \leq N\}|.$$
Given two parameters $\epsilon$ and $\delta$, a Count-Min Sketch is constructed as a two-dimensional counter table with 
$d = \lceil \ln(1/\delta) \rceil$ rows and $w = \lceil e/\epsilon \rceil$ columns. Initially, all the counters inside the sketch are set to $0$. 

There are two fundamental operations for a CMS; the first one is {\em insert}($x$) which updates
 internal sketch counters to process the items in the stream. 
To insert $x \in \mathcal{U}$, the counters ${\tt cms}[i][h_i(x)]$ are incremented for $1 \leq i \leq d$, i.e., a counter from each row is incremented where the column IDs are obtained from the hash values. 
Algorithm~\ref{alg:cms_construct} gives the pseudocode to sequentially process  ${\tt s}[.]$ of size $N$ and construct a CMS. 
 
The second operation for CMS is {\em query}($x$) to estimate the frequency of $x \in \mathcal{U}$ as $$f'_x = min_{1 \leq i \leq d}\{{\tt cms}[i][h_i(x)]\}.$$
\noindent With $d \times w$ memory, the sketch satisfies that
 $f_x \leq f'_x$ and $\Pr\left(f'_x \geq f_x + \epsilon N\right) \leq \delta.$ Hence, the error is additive and always one-sided. Furthermore, for $\epsilon$ and $\delta$ small enough, the  error is also bounded with high probability. Hence, especially for frequent items with large $f_x$,  the ratio of the estimation to the actual frequency approaches to one.  
 
 \renewcommand{\baselinestretch}{0.9}
 \begin{algorithm}[htbp]
	\small
  	\caption{\textsc{CMS-Construction}} 
  	\KwIn{  $\epsilon$: error factor, $\delta$: error probability \\
	 	  \hspace*{8ex}${\tt s}[.]$: a stream with $N$ elements from $n$ distinct elements \\ 
		  \hspace*{8ex}$h_i(.)$: pairwise independent hash functions where for \\ 
		  \hspace*{13ex}$1\leq i \leq d$,  $h_i$: $\mathcal{U} \rightarrow \{1,\cdots,w\}$ and $w = \lceil e/\epsilon \rceil$\\}
	 \KwOut{ ${\tt cms}[.][.]$: a $d \times w$ counter sketch where $d = \lceil 1/\delta \rceil$ \\
	 }
	 	\For{$i\leftarrow 1$ \KwTo $d$}{
			\For{$j\leftarrow 1$ \KwTo $w$}{
				${\tt cms}$[i][j] $ \leftarrow 0$
			}
		}
		\For{$i\leftarrow 1$ \KwTo $N$}{
			$x  \leftarrow s[i]$
			\For{$j\leftarrow 1$ \KwTo $d$}{
				 $col \leftarrow h_j(x)$\\
				${\tt cms}$[j][$col$] $ \leftarrow {\tt cms}$[j][$col$] $ +1$
			}
		}
	\label{alg:cms_construct}
\end{algorithm} 	
\renewcommand{\baselinestretch}{1}
\vspace*{-4ex}

\paragraph{Tabulation Hash:} CMS requires pairwise independent hash functions to provide the desired properties stated above. A separate hash function is used for each row of the CMS with a range equal 
to the range of columns. In this work, we use tabulation hashing~\cite{zobrist1970} which has been recently analyzed by Patrascu and Thorup et al.~\cite{patrascu2012,thorup2017} and shown to provide  strong  statistical guarantees despite of its simplicity. Furthermore, it is even as fast as the classic multiply-mod-prime scheme, i.e., $(ax + b) \bmod p$. 

Assuming each element in $\mathcal{U}$ is represented in 32 bits~(the hash function can also be used to hash 64-bit stream items~\cite{thorup2017}) and the desired output is also 32 bits, tabulation hashing works as follows: first a $4 \times 256$ table is generated and filled with random 32-bit values. Given a 32-bit input $x$, each character, i.e., 8-bit value, of $x$ is used as an index for the corresponding row. Hence, four 32-bit values, one from each row, are extracted from the table. The bitwise {\tt XOR} of these 32-bit values are returned as the hash value. 

\section{Merged Tabulation with a Single Table}\label{sec:tab}
Hashing the same item with different members of a hash family is a common technique in sketching applied to reduce the error of the estimation. One can use a single row for CMS, i.e., set $d = 1$ and answer the query by reporting the value of the counter corresponding to the hash value. However, using multiple rows reduces the probability of having large estimation errors. 

Although the auxiliary data used in tabulation hashing are small and can fit into a cache, the spatial locality of the accessed table elements, i.e., their distance in memory, is deteriorating since each access is performed to a different table row~(of length 256). A naive, cache-friendly rearrangement of the entries in the tables is also not possible for applications performing a single hash per item; the indices for each table row are obtained from adjacent chunks in the binary representation of the hashed item which are usually not correlated. Hence, there is no relation whatsoever among them to help us to fix the access pattern for all possible stream elements.

\begin{figure}[htbp]
\begin{minipage}[c]{0.65\textwidth}
\includegraphics[width=\textwidth]{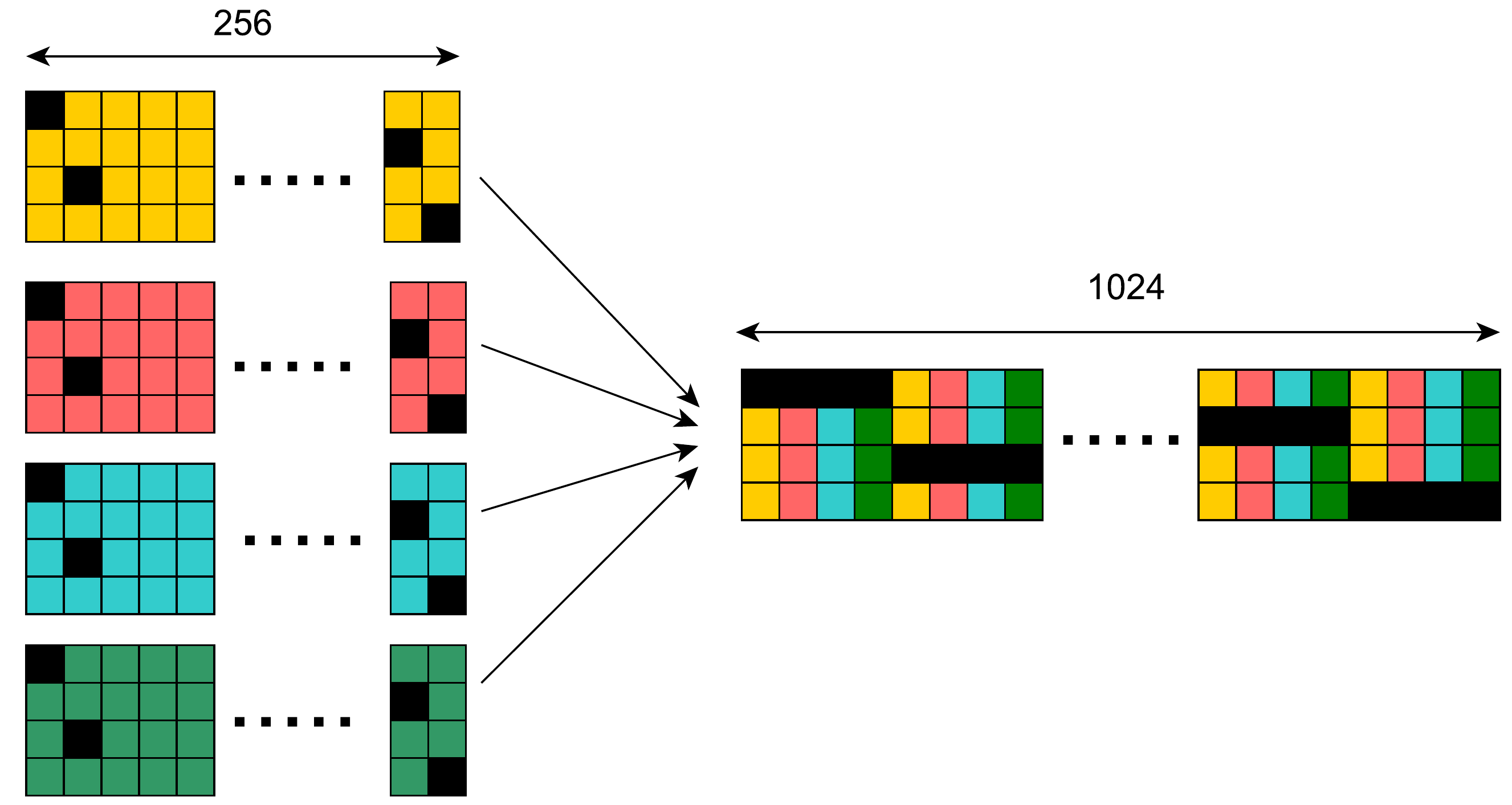}
  \end{minipage}\hfill\hfill\hfill
 \begin{minipage}[c]{0.28\textwidth}
\caption{\small{Memory access patterns for naive and merged tabulation for four hashes. The hash tables are colored with different colors. The accessed locations are shown in black.}}
\label{fig:merged_tabular_access}
  \end{minipage}
\end{figure}

For many sketches, the same item is hashed more than once. When tabulation hashing is used, this yields an interesting optimization; there exist multiple hash functions and hence, more than one hash table. Although, the entries in a single table is accessed in a somehow irregular fashion, the accessed coordinates in all the tables are the same for different tables as can be observed on the left side of Figure~\ref{fig:merged_tabular_access}. Hence, the columns of the tables can be combined in an alternating fashion as shown in the right side of the figure. In this approach, when only a single thread is responsible from computing the hash values for a single item to CMS, the cache can be utilized in a better way since the memory locations accessed by that thread are adjacent. Hence, the computation will pay the penalty for a cache-miss only once for each 8-bit character of a 32-bit item. This proposed scheme is called {\em merged tabulation}.

\section{Parallel Count-Min Sketch Construction}\label{sec:par}

Since multiple CMS sketches can be combined, on a multicore hardware, each thread can process a different part of the data (with the same hash functions) to construct a partial CMS. These partial sketches can then be combined by adding the counter values in the same locations.
Although this approach has been already proposed in the literature and requires no synchronization,
the amount of the memory it requires increases with increasing number of threads. We included this {\em one sketch to one core} approach in the experiments as one of the baselines. 

Constructing a single CMS sketch in parallel is not a straightforward task. One can assign an item to a single thread and let it perform all the updates (i.e., increment operations) on CMS counters. The pseudocode of this parallel CMS construction is given in Algorithm~\ref{alg:cms_construct_par_nobuf}. However, to compute the counter values correctly, this approach requires a significant synchronization overhead; when a thread processes a single data item, it accesses an arbitrary column of each CMS row. Hence, race conditions may reduce the estimation accuracy. In addition, these memory accesses are probable causes of false sharing. 
To avoid the pitfalls stated above, one can allocate locks on the counters before every increment operation. However, such a   synchronization mechanism is too costly to be applied in practice. 

\renewcommand{\baselinestretch}{0.9}
 \begin{algorithm}[htbp]
  	\small
  	\caption{\textsc{Naive-Parallel-CMS}} 
  	\SetAlgoNoLine
  	\KwIn{  $\epsilon$: error factor, $\delta$: error probability \\
	 	  \hspace*{7ex}${\tt s}[.]$: a stream with $N$ elements from $n$ distinct elements \\ 
		  \hspace*{7ex}$h_i(.)$: pairwise independent hash functions where for \\ 
		  \hspace*{13ex}$1\leq i \leq d$,  $h_i$: $\mathcal{U} \rightarrow \{1,\cdots,w\}$ and $w = \lceil e/\epsilon \rceil$\\
		  \hspace*{7ex}$\tau$: no threads\\ }
	 \KwOut{ ${\tt cms}[.][.]$: a $d \times w$ counter sketch where $d = \lceil 1/\delta \rceil$ \\
	 }
		Reset all the ${\tt cms}[.][.]$ counters to 0 (as in Algorithm~\ref{alg:cms_construct}).\\
		\For{$i\leftarrow 1$ \KwTo $N$ {\bf in parallel}}{
			$x  \leftarrow s[i]$\\
			${\tt hashes}[.] \leftarrow$ {\sc MergedHash}($x$)

			\For{$j\leftarrow 1$ \KwTo $d$}{
				$col \leftarrow {\tt hashes}[j] $ \\
				${\tt cms}$[j][$col$] $ \leftarrow {\tt cms}$[j][$col$] $ +1$ (\em{must be a critical update})
			}
		}
	\label{alg:cms_construct_par_nobuf}
\end{algorithm} 	
\renewcommand{\baselinestretch}{1}

In this work, we propose a {\em buffered parallel} execution to alleviate the above mentioned issues; we (1) divide the data into batches and (2) process a single batch in parallel in two phases; a) merged-hashing and b) CMS counter updates. In the proposed approach, the threads synchronize after each batch and process the next one. For batches with $b$ elements, the first phase requires a buffer of size $b \times d$ to store the hash values, i.e., column ids, which then will be used in the second phase to update corresponding CMS counters. Such a buffer allows us to use merged tabulation effectively during the first phase. In our implementation, the counters in a row are updated by the same thread hence, there will be no race conditions and probably much less false sharing. Algorithm~\ref{alg:cms_construct_par} gives the pseudocode of the proposed buffered CMS construction approach.\looseness=-1 

\renewcommand{\baselinestretch}{0.9}
 \begin{algorithm}[htbp]
  	\small
  	\caption{\textsc{Buffered-Parallel-CMS}} 
  	\KwIn{  $\epsilon$: error factor, $\delta$: error probability \\
	 	  \hspace*{7ex}${\tt s}[.]$: a stream with $N$ elements from $n$ distinct elements \\ 
		  \hspace*{7ex}$h_i(.)$: pairwise independent hash functions where for \\ 
		  \hspace*{13ex}$1\leq i \leq d$,  $h_i$: $\mathcal{U} \rightarrow \{1,\cdots,w\}$ and $w = \lceil e/\epsilon \rceil$\\
		  \hspace*{7ex}$b$: batch size (assumption: divides $N$)\\
		  \hspace*{7ex}$\tau$: no threads (assumption: divides $d$)\\ }
	 \KwOut{ ${\tt cms}[.][.]$: a $d \times w$ counter sketch where $d = \lceil 1/\delta \rceil$ \\
	 }
		Reset all the ${\tt cms}[.][.]$ counters to 0 (as in Algorithm~\ref{alg:cms_construct})\\[5pt]

		\For{$i\leftarrow 1$ \KwTo $N/b$}{
			$j_{end} \leftarrow i \times b$	 
			$j_{start} \leftarrow j_{end} - b + 1$
			
			\For{$j\leftarrow j_{start}$ \KwTo $j_{end}$ {\bf in parallel}}{
				$x  \leftarrow {\tt s}[j]$\\
				$\ell_{end} \leftarrow j \times d$\\
				$\ell_{start} \leftarrow \ell_{end} - d + 1$\\
				${\tt buf}[\ell_{start}, \cdots, \ell_{end}] \leftarrow$ {\sc MergedHash}($x$)
			}
			
			{{\bf Synchronize} the threads, e.g., with a {\em barrier}}
			
			\For{$t_{id}\leftarrow 1$ \KwTo $\tau$ {\bf in parallel}}{
				\For{$j\leftarrow 1$ \KwTo $b$ }{
					$nrows \leftarrow d / \tau$\\
					$r_{end} \leftarrow t_{id} \times nrows$\\
					$r_{start} \leftarrow r_{end} - nrows + 1$\\
					\For{$r\leftarrow r_{start}$ \KwTo $r_{end}$ }{
						$col \leftarrow {\tt buf}[((j-1) \times d) + r]$\\
						${\tt cms}[r][col] \leftarrow {\tt cms}[r][col] +1$
			}
			}
			}

			}
	\label{alg:cms_construct_par}
\end{algorithm} 	
\renewcommand{\baselinestretch}{1}

\section{Managing Heterogeneous Cores} \label{sec:load}

A recent trend on SBC design is heterogeneous multiprocessing which had been widely adopted by mobile devices. Recently, some ARM-based devices including SBCs use the {\em big.LITTLE} architecture equipped with power hungry but faster cores, as well as battery-saving but slower cores. The faster cores are suitable for compute-intensive, time-critical tasks where the slower ones perform the rest of the tasks and save more energy. In addition, tasks can be dynamically swapped between these cores on the fly. One of the SBCs we experiment in this study has an 8-core Exynos 5422 Cortex processor having four fast and four relatively slow cores. 

Assume that we have $d$ rows in CMS and $d$ cores on the processor; when the cores are homogeneous, Algorithm~\ref{alg:cms_construct_par} works efficiently with static scheduling since, each thread performs the same amount of merged hashes and counter updates. When the cores are heterogeneous, the first inner loop (for merged hashing) can be dynamically  scheduled: that ia a batch can be divided into smaller, independent chunks and the faster cores can hash more chunks. 
However, the same technique is not applicable to the (more time consuming) second inner loop where the counter updates are performed: in the proposed buffered approach, Algorithm~\ref{alg:cms_construct_par} divides the workload among the threads by assigning each row to a different one. When the fast cores are done with the updates, the slow cores will still be working. Furthermore, faster cores cannot help to the slower ones by stealing a portion of their remaining jobs since when two threads work on the same CMS row, race conditions will increase the error.

\begin{figure}[htbp]
\begin{minipage}[c]{0.65\textwidth}
\includegraphics[width=\textwidth]{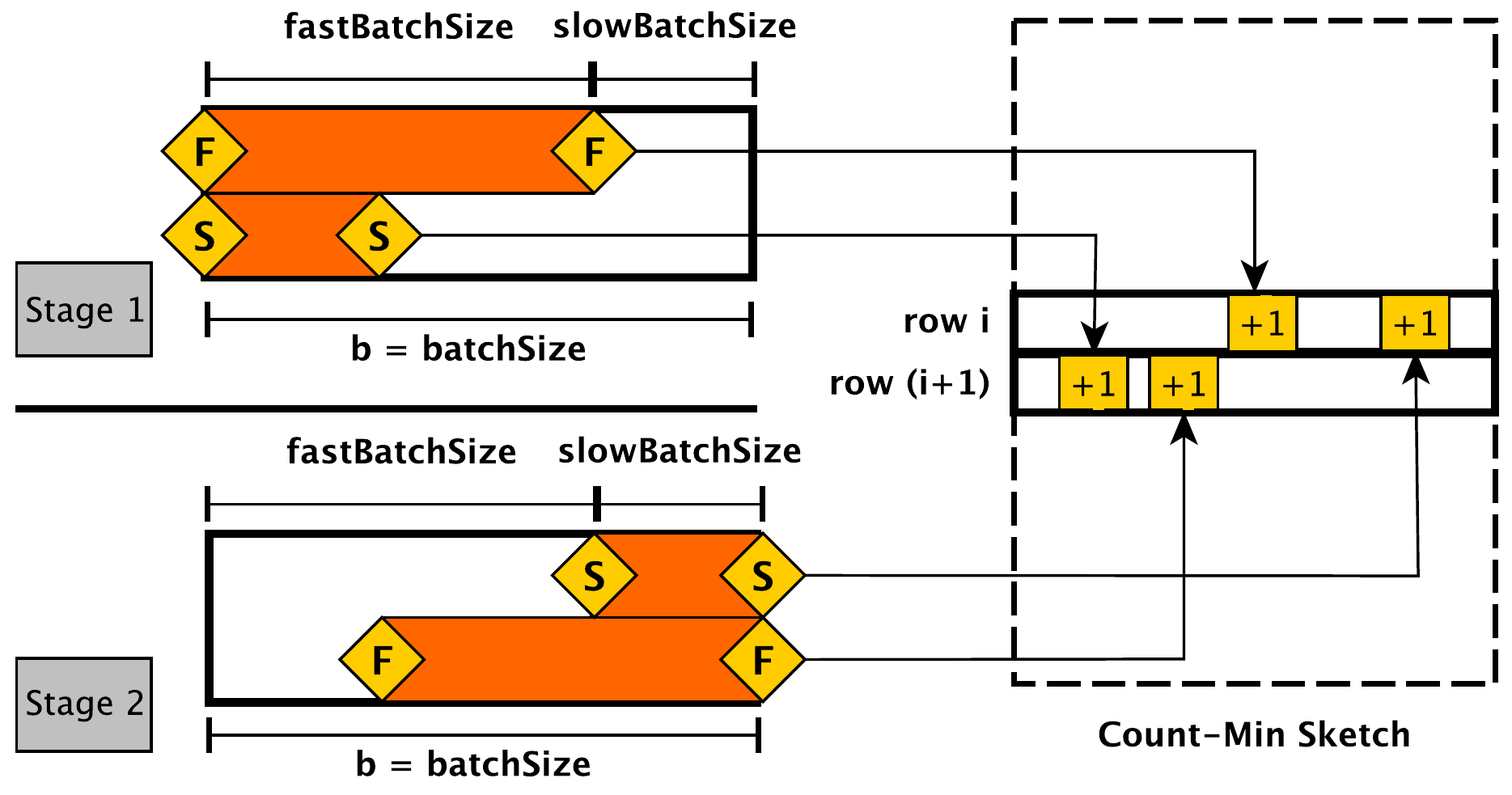}
  \end{minipage}\hfill\hfill\hfill
 \begin{minipage}[c]{0.32\textwidth}
\caption{\small{For a single batch, rows $i$ and $i+1$ of CMS are updated by a fast and a slow core pair in two stages. In the first stage, the fast core performs row $i$ updates and the slow core processes row $i+1$ updates. In the second stage, they exchange the rows and complete the remaining updates on the counters for the current batch.}}
\label{fig:fastslow}
  \end{minipage}
  \vspace*{-4ex}
\end{figure}

To alleviate these problems, we propose to pair a slow core with a fast one and make them update two rows in an alternating fashion. The batch is processed in two stages as shown in Figure~\ref{fig:fastslow}; in the first stage, the items on the batch are processed in a way that the threads running on faster cores update the counters on even numbered CMS rows whereas the ones running on slower cores update the counters on odd numbered CMS rows. When the first stage is done, the thread/core pairs exchange their row ids and resume from the item their mate stopped in the first stage. In both stages, the faster threads process $fastBatchSize$ items and the slower ones process $slowBatchSize$ items where $b = fastBatchSize + slowBatchSize.$\vspace*{-2ex}

To avoid the overhead of dynamic scheduling and propose a generic solution, we start with $fastBatchSize = b/2$ and $slowBatchSize = b/2$ and by measuring the time spent by the cores, we dynamically adjust them to distribute the workload among all the cores as fairly as possible. Let $t_F$ and $t_S$ be the times spent by a fast and slow core, respectively, on average. Let $s_F = \frac{fastBatchSize}{t_F}$ and $s_S = \frac{slowBatchSize}{t_S}$ be the speed of these cores for the same operation, e.g., hashing, CMS update etc. We then solve the equation $\frac{fastBatchSize + x}{s_F} = \frac{slowBatchSize - x}{s_S}$ for $x$ and update the values as 
\begin{align*}
fastBatchSize &= fastBatchSize + x\\
slowBatchSize &= slowBatchSize - x
\end{align*}
for the next batch. One can apply this method iteratively for a few batches and use the average values to obtain a generic and dynamic solution for such computations. To observe the relative performances, we applied this technique both for hashing and counter update phases of the proposed buffered CMS generation algorithm.

\section{Experimental Results}\label{sec:exp}

We perform experiments on the following three architectures: 
\begin{itemize}[leftmargin=*]
\item {\bf Arch-1} is a server running on 64 bit CentOS 6.5 equipped with 64GB RAM and an Intel Xeon E7-4870 v2 clocked at 2.30 GHz and having 15 cores. Each core has a 32KB L1 and a 256KB L2 cache, and the size of L3 cache is 30MB. 
\item {\bf Arch-2} (Raspberry Pi 3 Model B+) is a quad-core 64-bit ARM Cortex A-53 clocked at 1.4 GHz equipped with 1 GB LPDDR2-900 SDRAM.  Each core has a 32KB L1 cache, and the shared L2 cache size is 512KB.
\item {\bf Arch-3} (Odroid XU4) is an octa-core heterogeneous multi-processor. There are four A15 cores running on 2Ghz and four A7 cores running on 1.4Ghz. The SBC is equipped with a 2GB LPDDR3 RAM. Each core has a 32KB L1 cache. The fast cores have a shared 2MB L2 cache and slow cores have a shared 512KB L2 cache. 
\end{itemize}
For  multicore parallelism, we use C++ and OpenMP. We use {\tt gcc 5.3.0} on {\bf Arch-1}. On {\bf Arch-2} and  {\bf Arch-3}, the {\tt gcc} version is {\tt 6.3.0} and {\tt 7.3.0}, respectively. For all architectures, {\tt -O3} optimization flag is also enabled.

To generate the datasets for experiments, we used {\em Zipfian} distribution~\cite{Zipf1935}. Many data in real world such as number of paper citations, file transfer sizes, word frequencies etc. fit to a Zipfian distribution with the shape parameter around $\alpha = 1$. Furthermore, the distribution is a common choice for the studies in the literature to benchmark the estimation accuracy of data sketches. To cover the real-life better, we used the shape parameter $\alpha \in \{1.1, 1.5\}$. Although they seem to be unrelated at first, an interesting outcome of our experiments is that the sketch generation performance depends not only the number of items but also the frequency distribution; when 
the frequent items become more dominant in the stream, some counters are touched much more than the others. This happens with increasing $\alpha$ and is expected to increase the performance since most of the times, the counters will already be in the cache. To see the other end of the spectrum, we also used {\em Uniform} distribution to measure the performance where all counters are expected to be touched the same number of times. 

We use $\epsilon \in \{10^{-3}, 10^{-4}, 10^{-5}\}$ and $\delta = 0.003$ to generate small, medium and large $d \times w$ sketches  where the number of columns is chosen as the first prime after $2/\epsilon$. Hence, the sketches have $w = \{2003, 20071, 200003\}$ columns and $d = \lceil \log_2(1/\delta) \rceil = 8$ rows. For the experiments on {\bf Arch-1}, we choose $N = 2^{30}$ elements from a universal set $\mathcal{U}$ of cardinality $n = 2^{25}$. For {\bf Arch-2} and   {\bf Arch-3}, we use $N = 2^{25}$ and $n = 2^{20}$. For all architectures, we used $b = 1024$ as the batch size. Each data point in the tables and charts given below is obtained by averaging ten runs. 

\subsection{Multi Table vs. Single Table}

\begin{figure}[htbp]
    \centering
    \begin{subfigure}[b]{0.47\textwidth}
        \centering
        \includegraphics[width = 6cm]{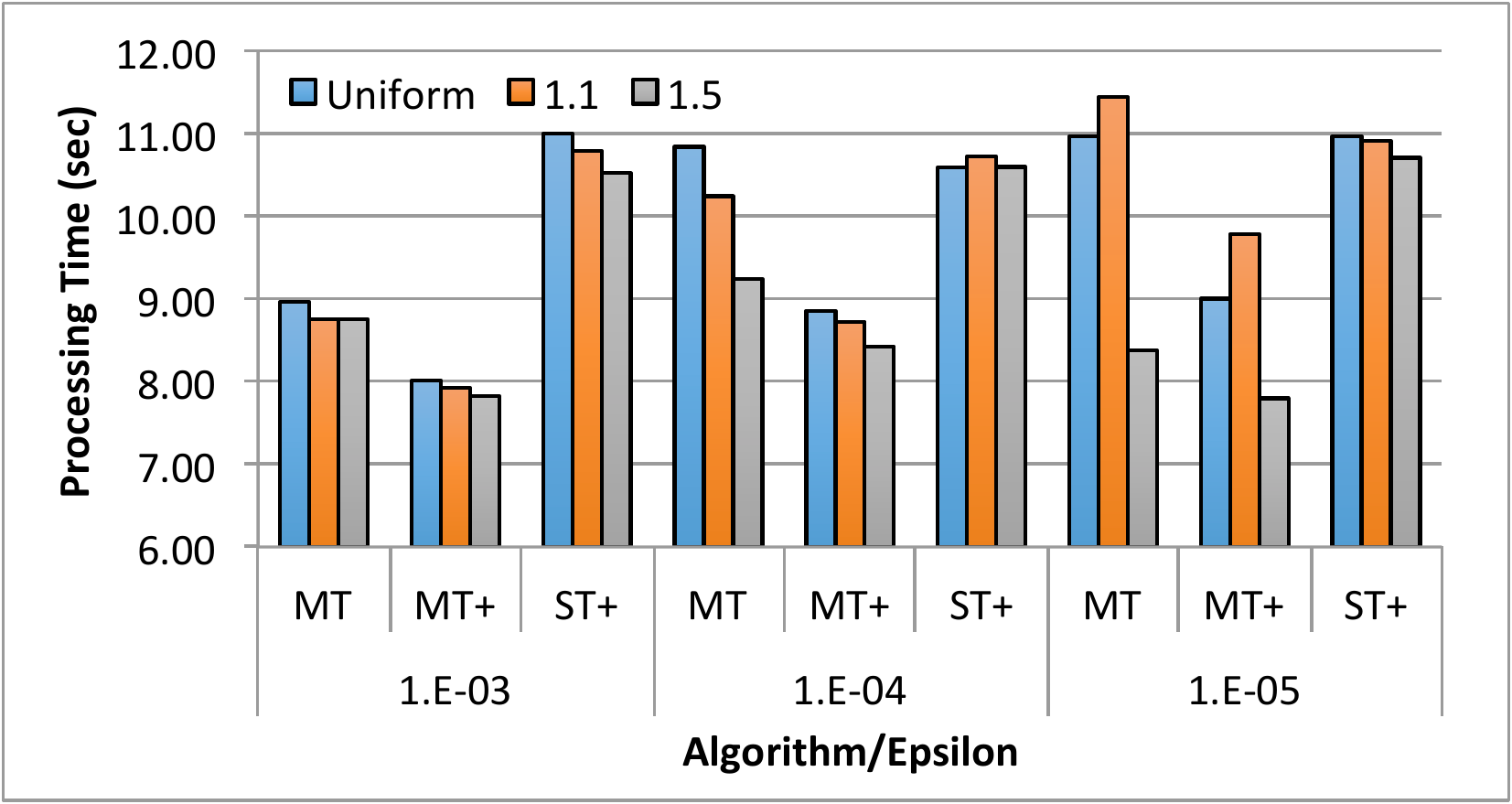}
        \caption{{\bf Arch-1} - 8 cores}
    \end{subfigure}\hspace*{1ex}
    ~
    \begin{subfigure}[b]{0.47\textwidth}
        \centering
        \includegraphics[width = 5.7cm]{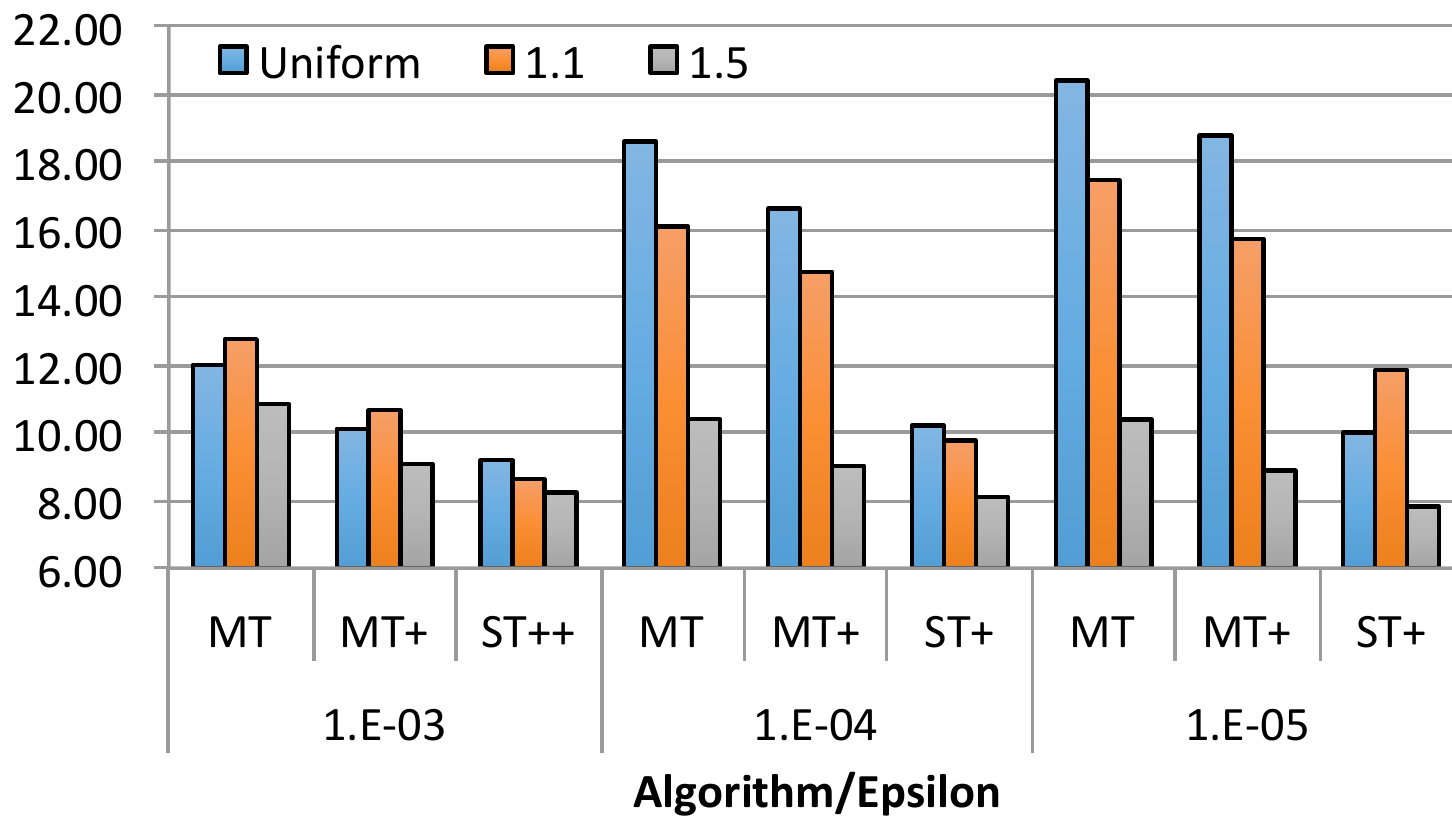}
        \caption{{\bf Arch-2} - 4 cores}
    \end{subfigure}
    
    \begin{subfigure}[b]{0.47\textwidth}
        \centering
        \includegraphics[width = 6.1cm]{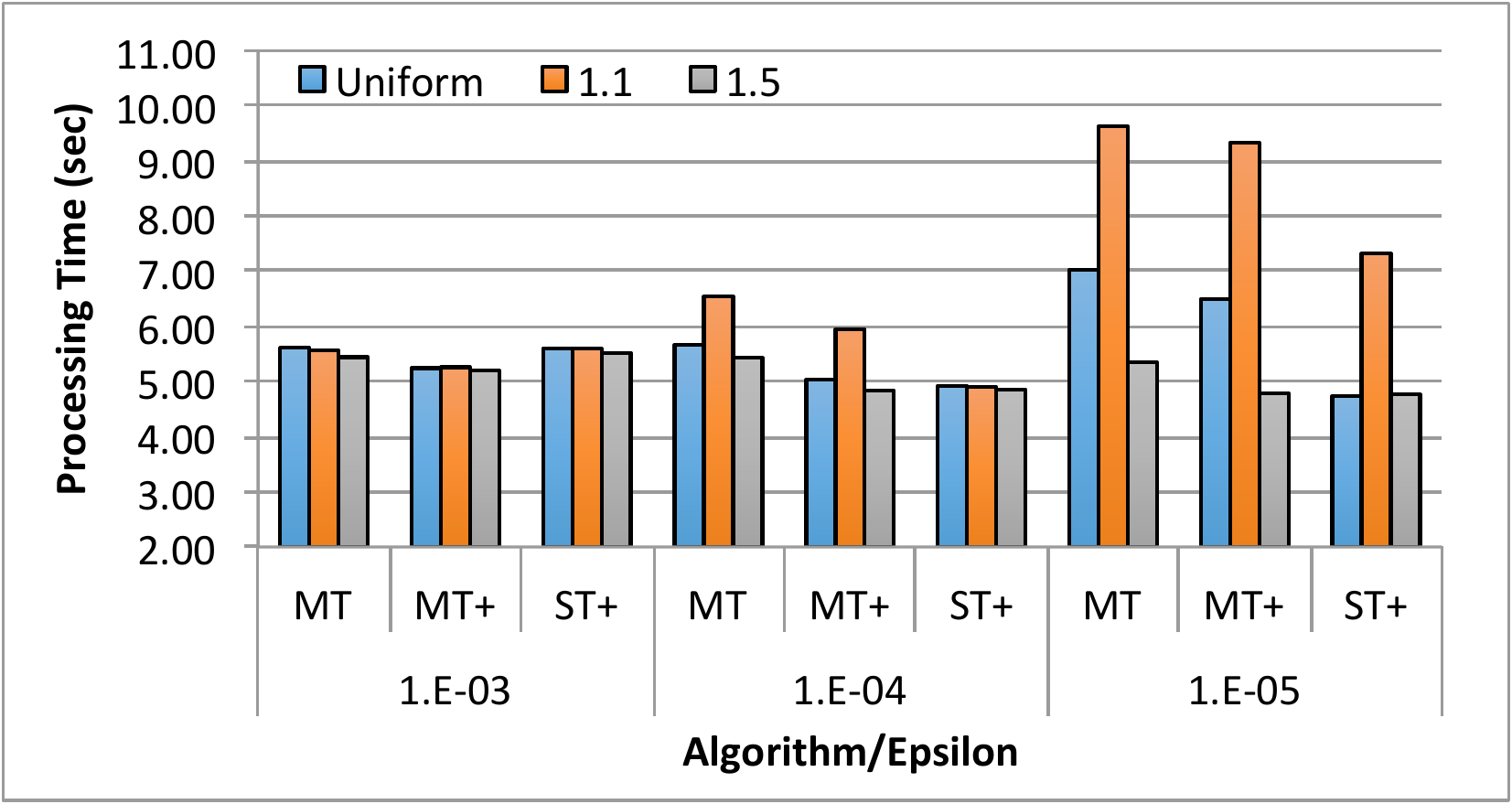}
        \caption{{\bf Arch-3} - 4 cores}
    \end{subfigure}\hspace*{1ex}
    ~
    \begin{subfigure}[b]{0.47\textwidth}
        \centering
        \includegraphics[width = 5.7cm]{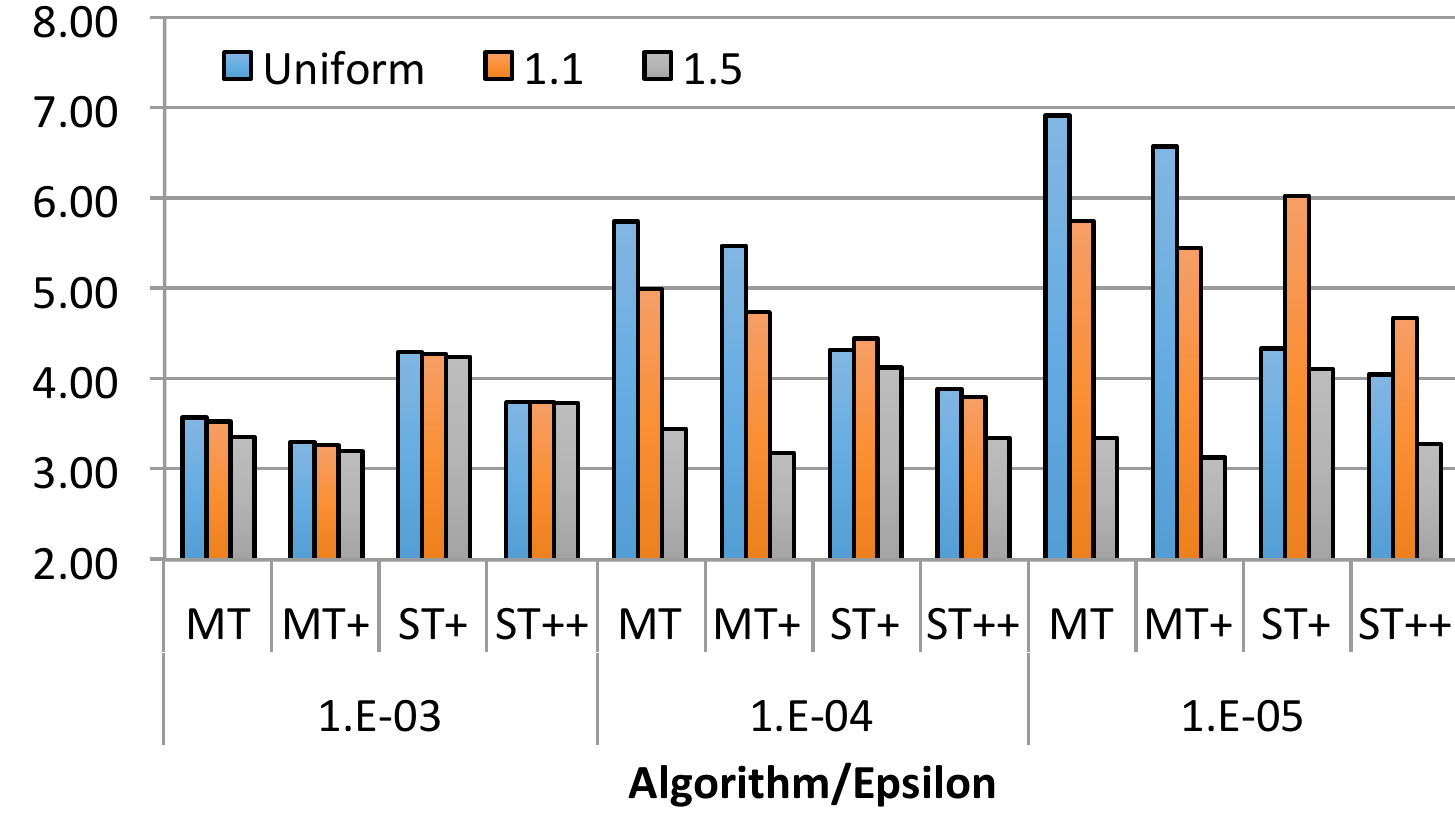}
        \caption{{\bf Arch-3} - 8 cores}
    \end{subfigure}
     \renewcommand{\baselinestretch}{0.93}
    \caption{\small{Performance comparison for multi-table~(MT) and single table~(ST) approaches. MT uses the one-sketch-per-core approach as suggested in the literature, MT+ is the MT-variant with merged tabulation. In all the figures, ST+ is the proposed scheme~(as in Algorithm~\ref{alg:cms_construct_par}), where in the last figure, ST++ is the ST+ variant using the  load-balancing scheme for heterogeneous cores as described in Section~\ref{sec:load}. For all the figures, the $x$-axis shows the algorithm and $\epsilon \in \{10^{-3}, 10^{-4}, 10^{-5}\}$ pair. The $y$-axis shows the runtimes in seconds; it does not start from 0 for a better visibility of performance differences. The first bar of each group shows the case when the data is generated using uniform distribution. The second and the thirds bars show the case for Zipfian distribution with the shape parameter $\alpha = 1.1$ and $1.5$, respectively.}}
     \renewcommand{\baselinestretch}{1}

    \label{fig:main}
\end{figure}

Although {\em one-sketch-per-core} parallelization, i.e., using partial, multiple sketches, is straightforward, it may not be a good approach for memory/cache restricted devices such as SBCs. The memory/cache space might be required by other applications running on the same hardware and/or other types of skeches being maintained at the same time for the same or a different data stream. 
Overall, this approach uses $(d \times w \times \tau)$ counters
where each counter can have a value as large as $N$; i.e., the memory consumption is
$(d \times w \times \tau \times \log N)$ bits. On the other hand, a single sketch with buffering
consumes $$(d \times ((w \times \log N) + (b \times \log w)))$$ bits since there are $(d \times b)$ entries in the
buffer and each entry is a column ID on CMS. For instance, with $\tau = 8$ threads, $\epsilon = 0.001$ and $\delta = 0.003)$, the one-sketch-per-core approach requires $(8 \times 2003 \times 8 \times 30) =$ 3.85Mbits whereas using single sketch requires
$(8 \times ((2003 \times 30) + (1024 \times 11))) =$ 0.57Mbits. Hence, in terms of memory footprint, using a single table pays off well. Figure~\ref{fig:main} shows the case for execution time.

In Figure~\ref{fig:main}.(a), the performance of the single-table~(ST+) and multi-table~(MT and MT+) approaches are presented on {\bf Arch-1}. Although ST+ uses much less memory, its performance is not good due to all the time spent while buffering and synchronization. The last level cache size on  {\bf Arch-1} is 30MB; considering the largest sketch we have is 6.4MB~(with 4-byte counters), {\bf Arch-1} does not suffer from its cache size and MT+ indeed performs much better than ST+. However, as Fig.~\ref{fig:main}.(b) shows for {\bf Arch-2}, with a $512$KB last-level cache, the proposed technique significantly improves the performance, and while doing that, it uses significantly much less memory. As Fig.~\ref{fig:main}.(c) shows, a similar performance improvement on {\bf Arch-3} is also visible for medium~(640KB) and especially large~(6.4MB) sketches when only the fast cores with a 2MB last-level cache are used.\looseness=-1 

Figure~\ref{fig:main} shows that the performance of the algorithms vary with respect to the distribution. As mentioned above, the variance on the frequencies increases with increasing $\alpha$. For uniform and Zipfian($1.1$), the execution times tend to increase with sketch sizes. Nevertheless, for $\alpha = 1.5$, sketch size does not have a huge impact on the performance, since only the {\em hot} counters of the most frequent items are frequently updated. Although each counter has the same chance to be a hot counter, the effective sketch size reduces significantly especially for large sketches. This is also why the runtimes for many configurations are less for $\alpha = 1.5$.\looseness=-1 

\begin{figure*}[htbp]
	\begin{subfigure}[t]{0.49\textwidth}
		\includegraphics[width=\linewidth]{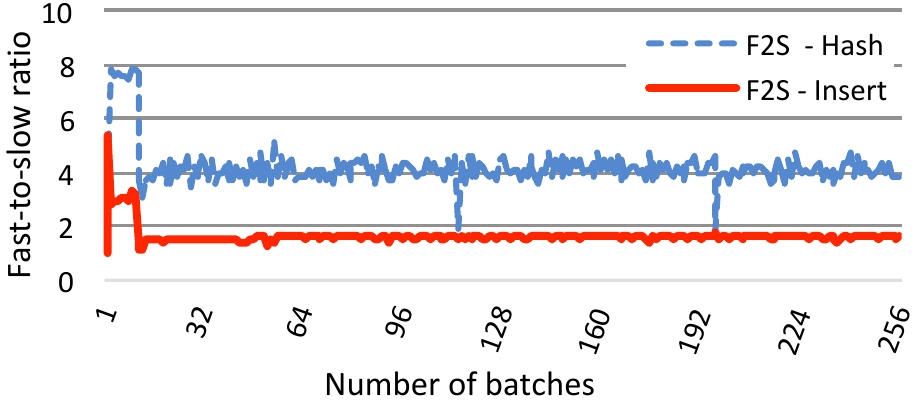}
		\caption{{\bf small} on {\bf Arch-3}}
		\label{fig:fs-small}
	\end{subfigure}\hspace*{3ex}
	\begin{subfigure}[t]{0.49\textwidth}
		\includegraphics[width=\linewidth]{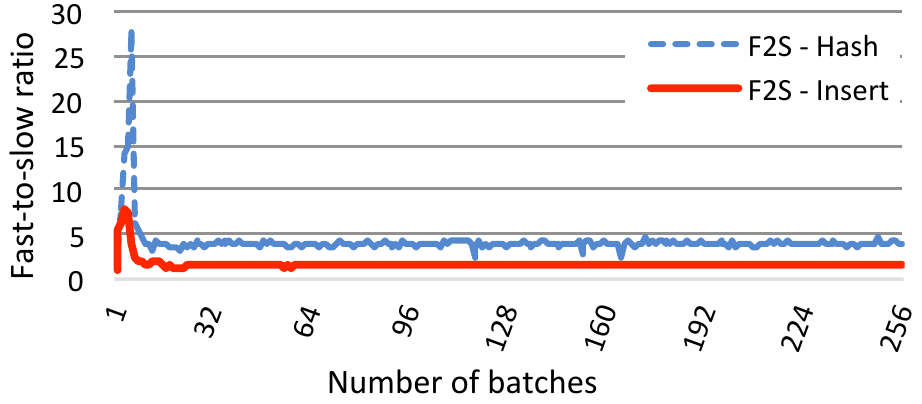}
		\caption{{\bf medium} on {\bf Arch-3}}
		\label{fig:fs-large}
	\end{subfigure}
	
	\caption{\small{Plots of fast-to-slow ratio $F2S = \frac{fastBatchSize}{slowBatchSize}$ of hashing and CMS update phases for consecutive batches and for small~(left) and medum~(right) sketches.}}
	\label{fig:load}
	\vspace*{-6ex}
\end{figure*}

\subsection{Managing Heterogeneous Cores} 

To utilize the heterogeneous cores on {\bf Arch-3}, we applied the smart load distribution described in Section~\ref{sec:load}. We pair each slow core with a fast one, virtually divide each batch into two parts, and make the slow core always run on smaller part. 
As mentioned before, for each batch, we dynamically adjust the load distribution based on the previous runtimes. Figure~\ref{fig:load} shows the ratio $F2S = \frac{fastBatchSize}{slowBatchSize}$ for the first $256$ batches of small and medium sketches. The best F2S changes w.r.t. the computation performed; for hashing, a 4-to-1 division of workload yields a balanced distribution. However, for CMS updates, a 1.8-to-1 division is the best. As the figure shows, the F2S ratio becomes stable after a few batches for both phases. Hence, one can stop the update process after $\thicksim$$30$  batches and use a constant F2S for the later ones. As Fig.~\ref{fig:main}.(d) shows, ST++, the single-table approach both with merged tabulation and load balancing, is always better than ST+. Furthermore, when $\tau = 8$, with the small 512KB last-level cache for slower cores, the ST++ improves MT+ much better~(e.g., when the medium sketch performance in Figs.~\ref{fig:main}.(c) and~\ref{fig:main}.(d) are compared).  Overall, smart load distribution increases the efficiency by $15\%$--$30\%$ for $\tau = 8$ threads.\looseness=1  

\subsection{Single Table vs. Single Table}

For completeness, we compare the performance of the proposed single-table approach. i.e., ST+ and ST++, with that of Algorithm~\ref{alg:cms_construct_par_nobuf}. However, we observed that using {\em atomic} updates drastically reduces its performance. Hence, we use the algorithm in a {\em relaxed} form, i.e., with non-atomic updates. Note that in this form, the estimations can be different than the CMS due to race conditions. As Table~\ref{tab:throughputs} shows, with a single thread, the algorithms perform almost the same except for {\bf Arch-1} for which Alg~\ref{alg:cms_construct_par_nobuf} is faster. However, when the number of threads is set to number of cores, the proposed algorithm is much better due to the negative impact of false sharing generated by concurrent updates on the same cache line. In its current form, the proposed algorithm can process approximately 60M, 4M, and 9M items on  {\bf Arch-1},  {\bf Arch-2} and  {\bf Arch-3}, respectively. \looseness=-1     
\vspace{-2ex}

\begin{table}
\begin{center}
     \scalebox{0.90}{
       \begin{tabular}{r|rr|rr||r|rr|rr}
Zipfian  & \multicolumn{2}{c|}{{\bf Alg  3}~(ST+ and ST++)} & \multicolumn{2}{c||}{{\bf Alg  2} - relaxed} &Zipfian  & \multicolumn{2}{c|}{{\bf Alg  3}~(ST+ and ST++)} & \multicolumn{2}{c}{{\bf Alg  2} - relaxed}\\
 $\alpha = 1.1$  & $\tau = 1$ & $\tau \in \{4,8\}$& $\tau = 1$  & $\tau \in \{4,8\}$ &  $\alpha = 1.5$  & $\tau = 1$ & $\tau \in \{4,8\}$& $\tau = 1$  & $\tau \in \{4,8\}$  \\\hline
{\bf Arch-1} &   17.6 & 60.0 & 22.6 & 17.8  &  {\bf Arch-1} & 17.9 & 57.6 & 22.6 & 12.9 \\
{\bf Arch-2} & 1.3 & 3.9 & 1.3 & 3.3 & \hspace*{2ex}  {\bf Arch-2} & 1.3 & 4.1 & 1.2 & 3.2 \\
{\bf Arch-3} &   1.6 & 9.0 & 1.6 & 6.6 &  {\bf Arch-3} & 1.6 & 9.0 & 1.7 & 6.1  \\
\end{tabular}
       }
       \caption{\small{Throughputs for sketch generation - million items per second. For each architecture, the number of threads is set to either one or the number of cores.}}
     \label{tab:throughputs}
     \end{center}
     \vspace{-11ex}
\end{table}

\section{Related Work}\label{sec:related}

CMS is proposed by Cormode and Muthukrishnan to summarize data streams~\cite{cormode2005}. Later, they comment on its parallelization~\cite{cormode2012} and briefly mention the single-table and multi-table approaches. There are studies in the literature employing synchronization primitives such as atomic operations for frequency counting~\cite{Das2009}. However, synchronization free approaches are more popular; Cafaro et al. propose an augmented frequency sketch for time-faded heavy hitters~\cite{cafaro2018}. They divided the stream into sub-streams and generated multiple sketches instead of a single one. A similar approach using multiple sketches is also taken by Mandal~et~al.~\cite{mandal18}.  CMS has also been used as an underlying structure to design advanced sketches. Recently, Roy et al. developed ASketch which filters high frequent items first and handles the remaining with a sketch such as CMS which they used for implementation~\cite{roy2016}. However, their parallelization also employs multiple filters/sketches. Another advanced sketch employing multiple CMSs for parallelization is FCM~\cite{Thomas2007}. 

Although other hash functions can also be used, we employ tabular hashing which is recently shown to provide good statistical properties and reported to be fast~\cite{thorup2017,Dahlgaard2017}. When multiple hashes on the same item are required, which is the case for many sketches, our merging technique will be useful for algorithms using tabular hashing. 

To the best of our knowledge, our work is the first cache-focused, synchronization-free, single-table CMS generation algorithm specifically tuned for limited-memory multicore architectures such as SBCs. Our techniques can also be employed for other table-based sketches such as Count Sketch~\cite{charikar2002} and CMS with conservative updates.

\section{Conclusion and Future Work}\label{sec:con}

In this work, we investigated the parallelization of Count-Min Sketch on SBCs. We proposed three main techniques: The first one, merged tabulation, is useful when a single is item needs to be hashed multiple times and can be used for different sketches. The second technique buffers the intermediate results to correctly synchronize the computation and regularize the memory accesses. The third one helps to utilize heterogeneous cores which is a recent trend on today's smaller devices. The experiments we performed show that the propose techniques improve the performance of CMS construction on multicore devices especially with smaller caches.

As a future work, we are planning to analyze the options on the SBCs to configure how much data/instruction cache they use, and how they handle coherency. We also want to extend the architecture spectrum with other accelerators such as FPGAs, GPUs, and more SBCs with different processor types. We believe that similar techniques we develop here can also be used for other sketches. 

 \renewcommand{\baselinestretch}{0.97}
 \bibliographystyle{splncs04}
 \bibliography{sketch}

\begin{thebibliography}{10}
\providecommand{\url}[1]{\texttt{#1}}
\providecommand{\urlprefix}{URL }
\providecommand{\doi}[1]{https://doi.org/#1}

\bibitem{alon1996}
Alon, N., Matias, Y., Szegedy, M.: The space complexity of approximating the
  frequency moments. In: Proceedings of the Twenty-eighth Annual ACM Symposium
  on Theory of Computing. pp. 20--29. STOC '96, ACM, New York, NY, USA (1996)

\bibitem{cafaro2018}
Cafaro, M., Pulimeno, M., Epicoco, I.: Parallel mining of time-faded heavy
  hitters. Expert Systems with Applications  \textbf{96},  115 -- 128 (2018)

\bibitem{charikar2002}
Charikar, M., Chen, K., Farach-Colton, M.: Finding frequent items in data
  streams. In: Proceedings of the 29th International Colloquium on Automata,
  Languages and Programming. pp. 693--703. ICALP '02, Springer-Verlag, Berlin,
  Heidelberg (2002)

\bibitem{cormode2003}
Cormode, G., Korn, F., Muthukrishnan, S., Srivastava, D.: Finding hierarchical
  heavy hitters in data streams. In: Proc. 29th Int. Conf. on Very Large
  Databases~(VLDB'03). pp. 464--475

\bibitem{cormode2012}
Cormode, G., Muthukrishnan, M.: Approximating data with the count-min sketch.
  IEEE Softw.  \textbf{29}(1),  64--69 (2012)

\bibitem{cormode2005}
Cormode, G., Muthukrishnan, S.: An improved data stream summary: the count-min
  sketch and its applications. Journal of Algorithms  \textbf{55}(1),  58 -- 75
  (2005)

\bibitem{Dahlgaard2017}
Dahlgaard, S., Knudsen, M.B.T., Thorup, M.: Practical hash functions for
  similarity estimation and dimensionality reduction. In: Advances in Neural
  Information Processing Systems (NIPS). pp. 6618--6628 (2017)

\bibitem{Das2009}
Das, S., Antony, S., Agrawal, D., El~Abbadi, A.: Thread cooperation in
  multicore architectures for frequency counting over multiple data streams.
  VLDB Endow.  \textbf{2}(1),  217--228 (Aug 2009)

\bibitem{dobra2002}
Dobra, A., Garofalakis, M., Gehrke, J., Rastogi, R.: Processing complex
  aggregate queries over data streams. In: Proceedings of the 2002 ACM SIGMOD
  International Conference on Management of Data. pp. 61--72. SIGMOD '02, ACM,
  New York, NY, USA (2002)

\bibitem{gilbert2002}
Gilbert, A.C., Kotidis, Y., Muthukrishnan, S., Strauss, M.J.: How to summarize
  the universe: Dynamic maintenance of quantiles. In: Proceedings of the 28th
  International Conference on Very Large Data Bases. pp. 454--465. VLDB '02,
  VLDB Endowment (2002)

\bibitem{mandal18}
Mandal, A., Jiang, H., Shrivastava, A., Sarkar, V.: Topkapi: Parallel and fast
  sketches for finding top-{K} frequent elements. In: NeurIPS 2018,
  Montr{\'{e}}al, Canada. pp. 10921--10931 (2018)

\bibitem{muthukrishnan2005}
Muthukrishnan, S.: Data streams: Algorithms and applications. Found. Trends
  Theor. Comput. Sci.  \textbf{1}(2),  117--236 (Aug 2005)

\bibitem{patrascu2012}
P\v{a}tra\c{s}cu, M., Thorup, M.: The power of simple tabulation hashing. J.
  ACM  \textbf{59}(3),  14:1--14:50 (Jun 2012)

\bibitem{roy2016}
Roy, P., Khan, A., Alonso, G.: Augmented sketch: Faster and more accurate
  stream processing. In: Proceedings of the 2016 International Conference on
  Management of Data. pp. 1449--1463. SIGMOD '16, ACM, New York, NY, USA (2016)

\bibitem{Thomas2007}
Thomas, D., Bordawekar, R., Aggarwal, C., Yu, P.S.: {A Frequency-aware Parallel
  Algorithm for Counting Stream Items on Multicore Processors}. Tech. rep., IBM
  (2007)

\bibitem{thorup2017}
Thorup, M.: Fast and powerful hashing using tabulation. Commun. ACM
  \textbf{60}(7),  94--101 (2017)

\bibitem{Zipf1935}
Zipf, G.: The Psychobiology of Language: An Introduction to Dynamic Philology.
  M.I.T. Press, Cambridge, Mass. (1935)

\bibitem{zobrist1970}
Zobrist, A.L.: A new hashing method with application for game playing.
  Technical Report 88  (University of Wisconsin, Madison, Wisconsin, 1970)

\end{thebibliography}
  \renewcommand{\baselinestretch}{1}

\end{document}